\documentstyle[12pt,epsf]{article}

\title{\ Linear Coasting in Cosmology and SNe Ia}
\author{Abha Dev\thanks{E--mail : abha@ducos.ernet.in},
Meetu Sethi\thanks{E--mail : meetu@ducos.ernet.in}, and
Daksh Lohiya\thanks{E--mail : dlohiya@ducos.ernet.in} \\
       {\em Department of Physics and Astrophysics,} \\
       {\em University of Delhi, Delhi-110007, India}
       }

\begin {document}
\baselineskip=2\baselineskip
\maketitle 

\baselineskip= 15pt

\vskip 1cm
\centerline{\bf Abstract}

	A strictly linear evolution of the cosmological expansion scale factor
is a characteristic feature in several classes of alternative gravity 
theories as  also  in the standard (big-bang) model with specially 
chosen equations 
of state of matter. Such an evolution has no free parameters as far as
the classical cosmological tests are concerned and should therefore
be easily falsifiable. In this article we demonstrate how such 
models present very good fits to the current supernovae 1a data. We discuss
the overall viability of such models. 

\vfil\eject

	Large scale homogeneity and isotropy observed in the universe
suggests the following [Friedman-Robertson-Walker (FRW)] form
for the spacetime metric:
\begin{equation}
\label{1}
ds^2 = dt^2 - a(t)^2[{dr^2\over {1 - kr^2}} + r^2(d\theta^2 
+ sin^2\theta d\phi^2)]
\end{equation}
Here $k = \pm 1, 0$ is the curvature constant.
In standard ``big-bang'' cosmology, the scale factor $a(t)$ is determined by
the equation of state of matter  and Einstein's equations. 
The form for the scale factor determines the response of a chosen 
model to the  three classical cosmological tests, viz: (1) The
galaxy number count as a function of redshift; (2) The angular diameter
of ``standard'' objects (galaxies) as a function of redshift; and finally
(3) The apparent luminosity of a ``standard candle'' as a function of
redshift. The first two tests are marred by evolutionary effects and for
this reason have  fallen into disfavour as reliable indicators of 
a viable model. However, the discovery of Supernovae type Ia [SNe Ia] 
as reliable 
standard candles  has raised hopes of elevating 
the status of the third test to that
of a precision measurement that could determine the viability 
of a cosmological model. The main reason for regarding these objects
as reliable standard candles are their large luminosity, small dispersion 
in their peak luminosity and a fairly accurate modeling of their 
evolutionary features. Recent measurements on 42 high redshift 
SNe Ia's reported
in the supernovae cosmology project 
\cite{perl} together with the observations of
the 16 lower redshift SNe Ia's of the Callan-Tollolo survey \cite{ham,ham1} 
have been used to determine the cosmological 
parameters $\Omega_{\Lambda}$ and $\Omega_{M}$. The data eliminates the 
``minimal  inflationary'' prediction defined by $\Omega_{\Lambda} = 0$ and 
$\Omega_{M} = 1$. The data can however, be used to assess a ``non-minimal 
inflationary cosmology'' defined by $\Omega_\Lambda \ne 0$,~
$\Omega_{\Lambda} +\Omega_{M} = 1$.
The maximum likelihood analysis following from such a study has yielded 
the values $\Omega_M = 0.28 \pm 0.1$ and $\Omega_\Lambda = 0.72 \pm 0.1$
\cite{perl1,perl2,wendy,branch}.

	In this article we explore the concordance of the SNe Ia data 
with a FRW cosmology in which the scale factor evolves linearly with time:
$a(t) \propto t$. The motivation for such an endeavour 
comes from several reasons.
First of all, such a cosmology does not suffer from the horizon problem. 
Horizons occur in models with $a(t) \approx t^\alpha$ for $\alpha < 1$. 
Secondly,  linear evolution of the scale factor is supported in alternative 
gravity theories (eg. non-minimally coupled scalar tensor theories)
where it turns out to be independent of the matter 
equation of state. The scale factor in such theories
does not constrain the matter density parameter thereby curing the 
flatness problem.  
Further the age estimate of the ($a(t) \propto t$) 
universe, deduced from a measurement of the Hubble parameter, is given
by $t_o = (H_o)^{-1}$. This is $\approx$ 50\% greater than the age 
inferred from the same measurement in standard (cold) matter dominated
cosmology (without the 
cosmological constant). This would make the age estimate comfortably concordant
with age estimates of old clusters. Finally, 
a linear coasting cosmology,
independent of the equation of state of matter, 
is a generic feature in a class of models that attempt to dynamically 
solve the Cosmological constant
problem \cite{wein,dol,ford}. Such models have a scalar 
field non-minimally coupled
to the large scale scalar curvature of the universe. 
With the evolution of time, the non-minimal coupling
diverges, the scale factor quickly approaches linearity and the non-minimally
coupled field acquires a stress energy that cancels the vacuum energy 
in the theory.  

	There have been other gravity models that also account for a linear
evolution of the scale factor. Notable among such models is a recent idea
of Allen \cite{allen} in which such a scaling 
results in an $SU(2)$ cosmological 
instanton dominated  universe. Yet another possibility derives from the Weyl
gravity theory of Manheim and 
Kazanas \cite{mann}. Here again the FRW scale factor
approaches a linear evolution at late times. 

	What makes the above ideas particularly appealing is a recent
demonstration \cite{annu} of  primordial nucleosynthesis 
not to be an impediment for a linear coasting cosmology. 
For the currently favoured
value of 65 km /sec /Mpc for the Hubble parameter, it follows that for
baryon entropy ratio
$\eta \approx 5\times 10^{-9}$ we can get just the 
right amount of Helium $23.8 \% $. Besides, one also gets a primordial
metallicity quite close to the lowest 
observed metallicities. The only problem that one has to contend with is 
the significantly low yields of deuterium in such a cosmology. However,
as pointed out in \cite{annu}, the amount of Helium produced is quite 
sensitive to $\eta$ in such models. In an inhomogeneous universe, therefore,
one can have the helium to hydrogen ratio to have a large variation. 
Deuterium can be produced by a spallation process much later in the history
of the universe \cite{eps}. If one considers spallation of a helium 
deficient cloud onto a helium rich cloud, it is easy to produce deuterium
as demonstrated by Epstein - but without overproduction of Lithium.
This result can be used to soften nucleosynthesis 
constraints on a general power law cosmology.

	Finally we must mention that within the framework of {\it standard}
cosmology there are rather strong constraints on the allowable equations of 
state. Perlmutter et al \cite{ptw} have demonstrated that  a general equation of
state: $P = w\rho$ in standard cosmology, is constrained to
$w \leq -2/3$. Linear coasting, possible  in standard cosmology for
$w = -1/3$ as also for empty models, would therefore appear to be disfavoured. 
However, as pointed out in \cite{perl}, such a linear coasting, though a bit 
improbable, is definitely not ruled out. Any mechanism for dimming the high redshift
SNe Ia by .15 magnitude would make linear coasting concordant. This is not at all 
discouraging - considering the fact that the intrinsic uncertainty in the luminosity for most high redshift
objects is in excess of .15. Further, as reported in this article, statistical 
measures of fit
such as the $\chi^2$ per degree of freedom for a linear coasting cosmology turn out
to be comparable to the best fits reported in \cite{perl} $\approx 1.17$. 
In any case the results presented in this article
would be of interest for alternative gravity models mentioned earlier.
	
\vskip 0.5cm
\centerline{\bf A coasting cosmology and Type Ia Supernovae}
\vskip 0.2cm

	 In this section we explore the concordance of a linear coasting
 cosmology with the recent extra-galactic Type Ia supernovae apparent
 magnitude - redshift data \cite{perl,perl1}.

The apparent magnitude of an object is related to the luminosity
distance $d_L$ by the
following relation,
\begin{equation}
m =  25 + M  + 5logd_L
\end{equation}
with $d_L$ defined as
\begin{equation}
d_L = a_o r (1 + z)
\end{equation}
Here r is the comoving radial coordinate.
In the standard model,
\begin{eqnarray}
D_L( z;\Omega_{M},\Omega_{\Lambda}) 
&\equiv& d_L H_o \\
&=& {c(1 + z) \over \sqrt{|\kappa|}}  \nonumber\\
&& {}\emph{S}(\sqrt{|\kappa|} \int_{0}^{z} {dz' \over \sqrt{[ ( 1 + z')^2( 1 + \Omega_{M} z') - z'(2 + z')\Omega_{\Lambda}]}}) 
\end{eqnarray}
where 
for $\Omega_{M} + \Omega_{\Lambda} > 1$, $\emph{S(x)} = sin(x)$ and 
$\kappa = 1 - \Omega_{M} - \Omega_{\Lambda}$;
for $\Omega_{M} + \Omega_{\Lambda} < 1$, $\emph{S(x)} = sinh(x)$ and
 $\kappa$ as above;
and for $\Omega_{M} + \Omega_{\Lambda} = 1,\emph{S(x)} = x$ and
$\kappa = 1$ \cite{perl2}.
In the low redshift limit, this equation reduces to the usual linear Hubble 
relation between m and $log(cz)$
\begin{equation}
m(z) = {\mathcal{M}} + 5 log(cz), 
\end{equation}
where $ {\mathcal{M}} = M - 5 log(H_{o}) + 25 $
This quantity can be measured from the apparent magnitude and redshift of low-
redshift samples of the standard candle independent of
$H_o$.
Using this value on recent measurements on 42 high redshift 
SNe Ia's reported
in the supernovae cosmology project \cite{perl} 
together with the observations of
the 16 lower redshift SNe Ia's of the Callan-Tollolo survey \cite{ham}, 
one can determine the cosmological 
parameters $\Omega_{\Lambda}$ and $\Omega_{M}$. The data eliminates the 
``minimal  inflationary'' prediction $\Omega_{\Lambda} = 0$, 
$\Omega_{M} = 1$ \cite{perl1,perl2,wendy,branch}.

     To explore the concordance of a linear coasting cosmology,
we consider a 
power law cosmology with the scale factor 
$a(t) = {\bar k} t^\alpha$, with ${\bar k},~\alpha$
arbitrary constants. The expression for the luminousity distance
is easily seen to be
\begin{equation}
d_L = ({\alpha \over H_{o}})^{\alpha}(1 + z)\emph{S}[{1 \over (1 -\alpha)} ({\alpha \over H_{o}})^{1 - \alpha}( 1 - (1 + z)^{ 1- {1 \over \alpha}})] 
\end{equation}
It is straightforward to discover 
the following relation between the apparent magnitude $m(z)$, the absolute
magnitude $M$  and the redshift $z$ of an object for such a cosmology:
\begin{equation}
m(z) = {\mathcal{M}} + 5logH_{o} + 5 log({\alpha \over H_o})^{\alpha}(1 + z) 
\emph{S}[{1 \over (1 -\alpha)}({\alpha \over H_o})^{1 - \alpha}
( 1 - (1 + z)^{ 1- {1 \over \alpha}})] 
\end{equation}
The above equations have simple forms in the the $\alpha \longrightarrow 1$
limit:
\begin{equation}
m(z) = {\mathcal{M}} + 5logH_{o} + 5 log[{{(1 + z)} \over H_o} 
\emph{S}(log(1 + z))]
\end{equation}
For example for an open $k = -1$, model
the above exact expression reduces to: 
\begin{equation}
  m(z) = 5 log(\frac{z^2}{2} + z) + \mathcal{M} 
\end{equation}

	Figure `1' \cite{perl} 
sums up  the  Supernova Cosmology project data 
for supernovae with redshifts
between 0.18 and 0.83 together with the set from the
Calan / Tololo  supernovae, at redshifts below 0.1. 

	To get an estimate of the goodness of concordance, 
we fitted the value of ${\mathcal{M}}$ using the 
low redshift data set \cite{ham}.
The value comes out to be ${\mathcal{M}} = -3.325$ and confirms the results of
Perlmutter et al \cite{perl}.

Next we used the 54 SN 1a data described by Perlmutter et al.
\cite{perl,perl1,perl2}
to obtain the 
value of the parameter $\alpha$ in eqn(8) by minimizing $\chi^2$:
\begin{equation}
\chi^2 = \Sigma {[m_B(z_i) - m_{Bi}]^2\over \sigma_{mBi}^2}
\end{equation}
The summation is over all the data points. Here $m_B(z_i)$ and
$m_{Bi}$  are respectively the theoretical and observed values for the 
apparent magnitude of a supernova and $\sigma_{mBi}$ the corresponding 
errors are displayed in tables I \& II of reference \cite{perl}.
For $\alpha = 1$ exactly, the limiting form for eqn(8), i.e. eqn(9)
is used.
The best fit turns out to be $\alpha = 1.001$, $k = -1$.
We used standard Monte-Carlo method to generate $10^4$ data sets
with a normal distribution around each data point consistent with the error 
bars for each of the supernovae. The best fit $\alpha$ for each of the data
sets was determined and the results are displayed in the 
histograms (Figures `3(a, b, c)'). Table I sums up the results for all 
$k = 0, \pm 1$ values. Figure 2 shows the hubble diagram corresponding to
 the best fit values of $\alpha$ for $ k = \pm 1,0$ in a power law cosmology.
  We find that linear coasting can not be eliminated 
for any of the models though the best fit is for the open model with
 $\alpha = 1.001 \pm{0.043}$, and $\chi^2_\nu$ (the minimum $\chi^2$ per
degree of freedom)  1.18. This is comparable to the corresponding values
(1.17) reported by Perlmutter et al for non-minimal inflationary 
cosmology parameter estimations. 
Thus a linear coasting is accommodated even in
the 68\% confidence region. This finds a passing mention in the analysis of
Perlmutter \cite{perl}
who noted that the curve for $(\Omega_\Lambda = \Omega_M = 0$ (for which
the scale factor would have a linear evolution) is ``practically identical to 
 $\bf{best fit}$ plot for an unconstrained cosmology''.

\begin{table}
\begin{center}
\begin{tabular}{|l|l|r||} \hline \hline
$k$   & $\alpha$ & $\chi^2_\nu$ \\ \hline 
-1 &  $ 1.004 \pm{0.043} $  & 1.179 \\ 
 0 &  $ 1.182 \pm{0.123} $  & 1.183 \\
 1 &  $ 0.990 \pm{0.099} $  & 1.397 \\ \hline
\end{tabular}
\caption{Summary of the Fit Results}
\end{center} 
\end{table}

\vfil
\eject 

\vskip 0.5 cm
{\bf Acknowledgments}

We thank Shiv K.Sethi and Tarun Deep Saini for many helpful discussions. 

\vskip 1cm

\bibliography{plain}

\begin {thebibliography}{99}
\bibitem{perl} S.Perlmutter, et al., {\emph astro-ph/9812133}.
\bibitem{perl1} S.Perlmutter, et al., \emph{Nature} {\bf 391}, 51 (1998). 
\bibitem{perl2} S.Perlmutter, et al., \emph{ApJ} {\bf 483}, 565 (1997).
\bibitem{ham} M.Hamuy, et al., \emph{Astron. J.} {\bf 112}, 2391 (1996).
\bibitem{ham1} M.Hamuy et al., \emph{Astron. J.} {\bf 109}, 1 (1995).
\bibitem{wendy} W.L. Freedman, J.R. Mould, R.C. Kennicutt, \&
B.F. Madore,{\emph astro-ph /9801080}.
\bibitem{branch} D.Branch, \emph{Ann.Rev.of Astronomy and Astrophysics}
{\bf 36}, 17 (1998), {\emph astro-ph/9801065}.
\bibitem{wein} S. Weinberg, \emph{Rev. Mod. Phys.} {\bf 61}, 1 (1989); 
\bibitem{dol} A. D. Dolgov in the 
{\it The Very Early Universe}, eds. G. Gibbons, S. Siklos, S. W. Hawking, 
C. U. Press, (1982);\emph{ Phys. Rev.} {\bf D55},5881 (1997).
\bibitem{ford} L.H. Ford, \emph {Phys Rev } {\bf D35},2339 (1987). 
\bibitem{allen} R.E.Allen \emph {astro-ph/9902042}. 
\bibitem{mann} P.Manheim \& D.Kazanas,\emph {Gen. Rel. \& Grav.} {\bf 22},289 
(1990).
\bibitem{annu}D.Lohiya, A. Batra, S. Mahajan, A. Mukherjee, {\emph nucl-th/
          9902022}, \emph{``Nucleosynthesis in a Simmering Universe''};
          \emph{Phys. Rev.} {\bf D60}, 108301 (2000)
\bibitem{eps} Epstein, R.I., Lattimer, J.M., and Schramm, D.N.
	 \emph{ Nature} {\bf 263}, 198 (1976).	
\bibitem{ptw}S.Perlmutter, M.S.Turner, M.White, \emph {Phys Rev, Let.} {\bf 83}, 670 (1999).
\bibitem{adam}A.Riess, et al.; \emph {Astron. J } {\bf 116}, 1009 (1998) 
\end {thebibliography}

\vfil
\eject

\begin{figure}[ht]
\vskip 15 truecm
\includegraphics{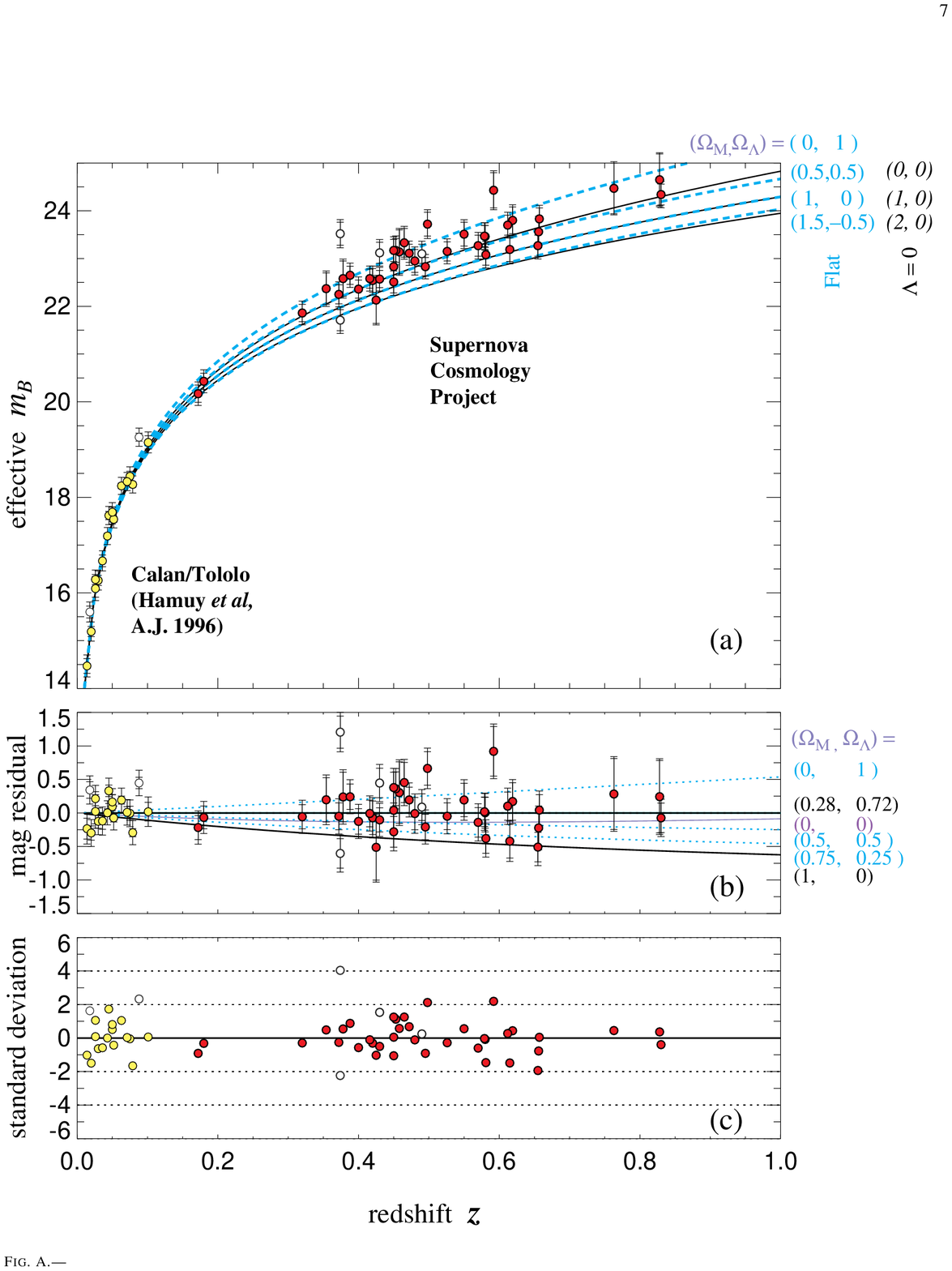}

\caption{  Hubble diagram, the magnitude residual and the uncertainty - normalized
residual plots taken from the supernova cosmology project. {\it ``The curve for
$(\Omega_\Lambda, \Omega_M) = (0,0)$ is practically identical to the best fit 
unconstrained cosmology''} \cite{perl}.}
\end{figure}

\begin{figure}[ht]
\vskip 15 truecm
\includegraphics{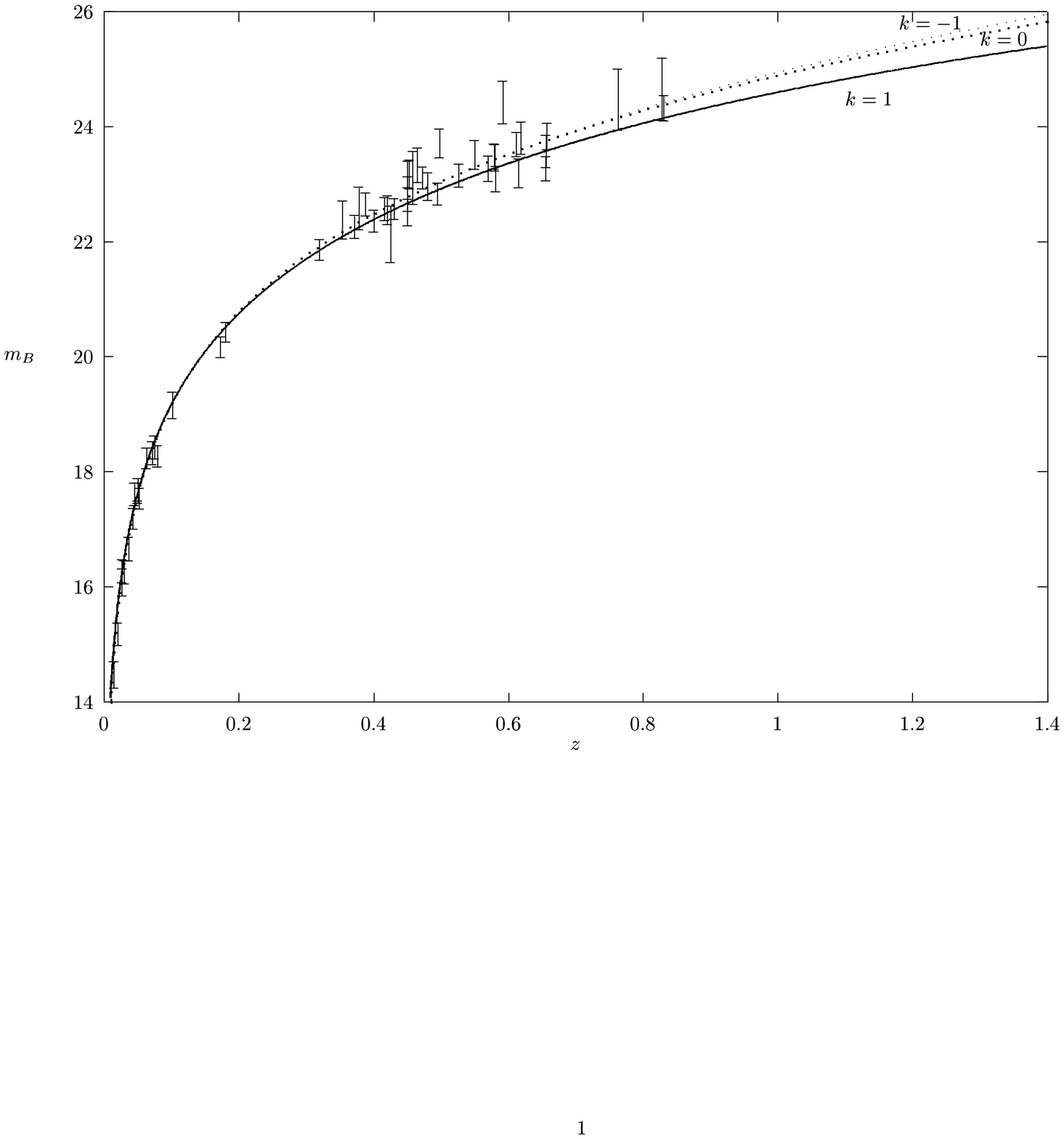}
\caption{  The Hubble diagram
for the best fit values of $\alpha$ for $k = \pm 1,0$ for a power law
cosmology projected against the data points of
 \cite{perl,perl1}}
\end{figure}

\pagebreak

\begin{figure}[ht]
\vskip 10truecm
\includegraphics{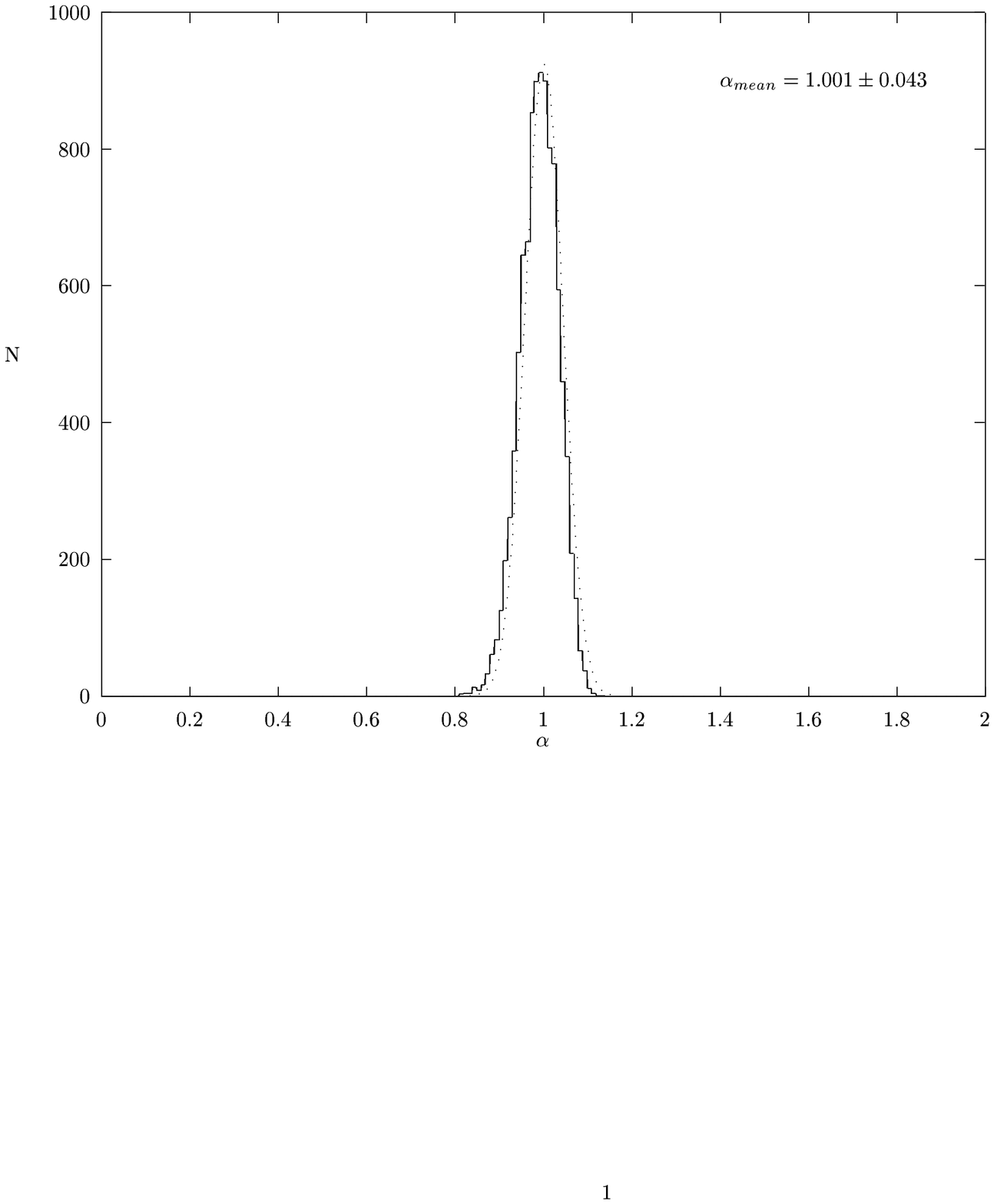}

\vskip 12cm

\includegraphics{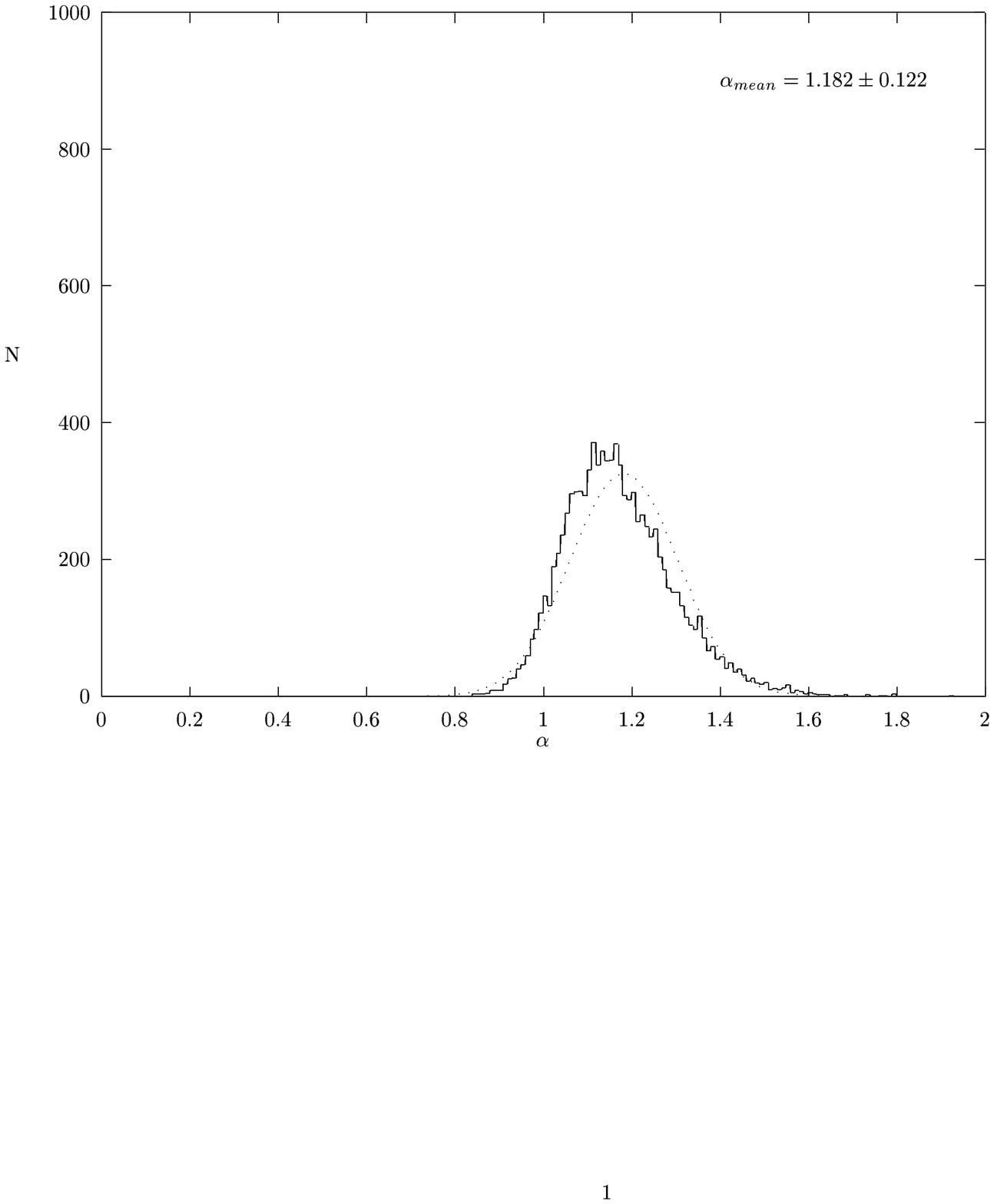}

\vskip 10cm
\end{figure}
\pagebreak
\begin{figure}[ht]
\vskip 10truecm
\includegraphics{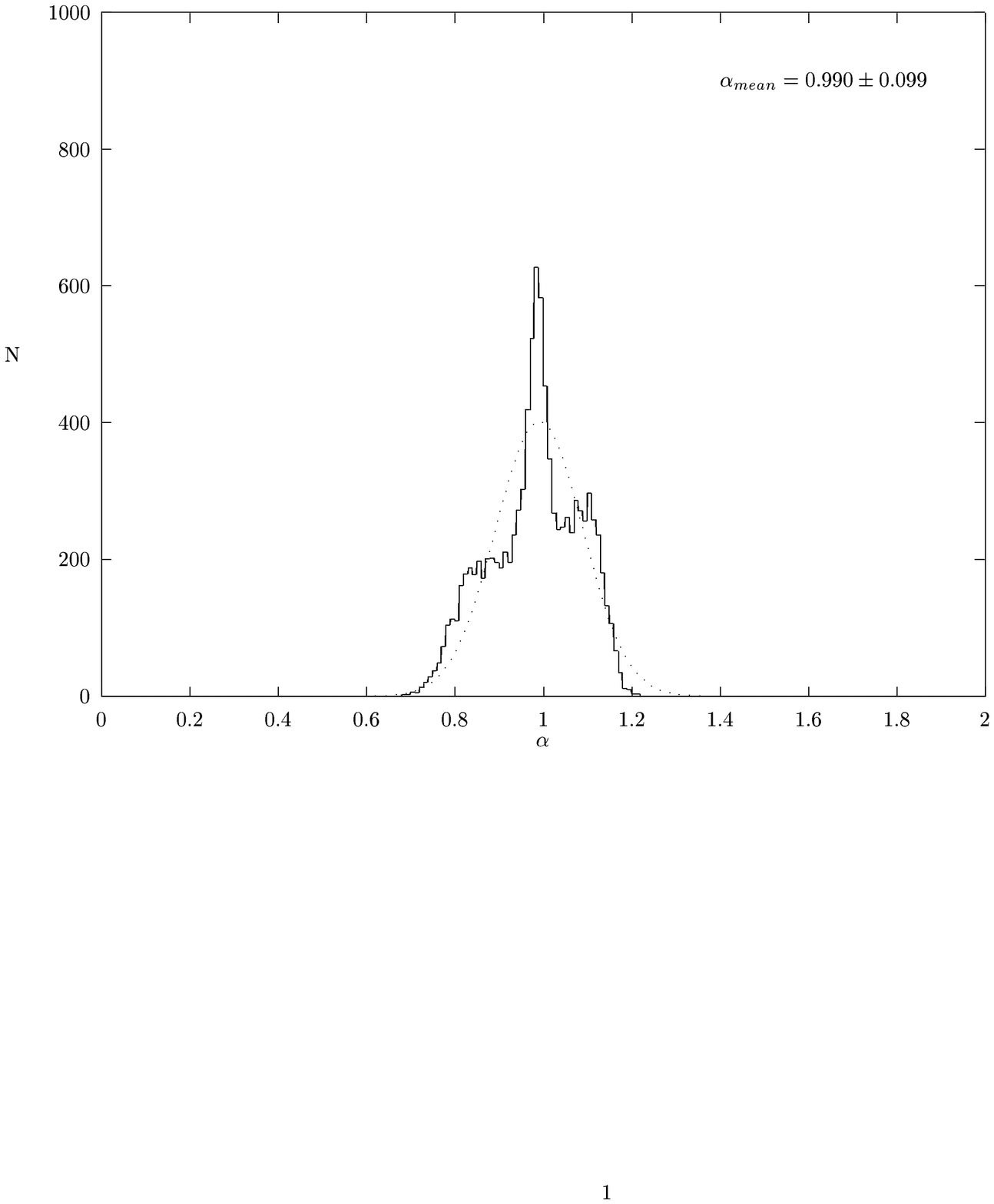}

\caption{  Histogram for $\alpha$ from $10^4$ data sets for  $k = -1,0,+1$
respectively.}
\end{figure}

\end{document}